\documentclass[twocolumn,showpacs,aps]{revtex4}

\usepackage{graphicx}% Include figure files

\begin{document}

%\draft %prints PACS numbers in

\title{Asymptotic construction of pulses in the discrete
        Hodgkin-Huxley model  for myelinated nerves}

\author{A. Carpio\cite{carpio:email} }

\affiliation{Departamento de Matem\'{a}tica Aplicada,
Universidad Complutense de Madrid, 28040 Madrid, Spain}

\date{ \today   }

\begin{abstract}

A quantitative description of pulses and wave trains in the spatially
discrete Hodgkin-Huxley model for myelinated nerves is given.
Predictions of  the shape and speed of the waves and the thresholds
for propagation failure are obtained. Our asymptotic predictions agree quite
well with numerical solutions of the  model and describe wave patterns
generated by repeated firing at a boundary.

\end{abstract}

%\pacs{05.45.-a,82.40.Ck,87.15.Aa,02.30.Ks}
\pacs{87.19.La,05.45.-a,82.40.Ck,02.30.Ks}

\maketitle

\section{Introduction}

\label{sec:intro}

Understanding wave propagation in discrete excitable media is
challenging because of poorly understood phenomena associated
with spatial discreteness \cite{sleeman,cbfhn,keener1,fath,cbn}.
The study of the transmission of nerve impulses along myelinated
axons is a paradigmatic example.
Myelinated nerve fibers, such as the motor axons of vertebrates,
are covered almost everywhere by a thick insulating coat of myelin.
Only a fraction of the active membrane is exposed, at small active
nodes called Ranvier nodes. The myelinated axons of motor
nerves can be very long, and contain hundreds or thousands of nodes
\cite{biophys}.
The wave activity jumps from one node to the next one giving rise
to ``saltatory''  propagation of impulses \cite{rushton}.
Saltatory conduction on myelinated nerve models has two
important features. One is the possibility of increasing the
speed of the nerve impulse while decreasing the diameter of the
nerve fiber \cite{scott1}. The other is propagation failure when
the myelin coat is damaged \cite{nature}, which causes diseases
such as multiple sclerosis.

The propagation of nerve impulses along a myelinated fiber can  be
described by the spatially discrete Hodgkin-Huxley system (HH)
\cite{scott1,keener2}. For typical experimental data, this system
couples equations for two fast variables and two slow variables. Most
analytical  studies of action potentials have focused
on discrete FitzHugh-Nagumo (FHN) models for one fast and one
slow variables \cite{kunov,erneux2,sleeman,hastings,cbfhn,tonnelier}
and  discrete bistable equations for the leading edge
\cite{richer,zinner,keener1,erneux1,fath,cbn}.
Careful computational studies of models including more biological
detail were carried out in \cite{fh,goldman,moore78}.

\begin{figure}
\begin{center}
\includegraphics[width=8cm]{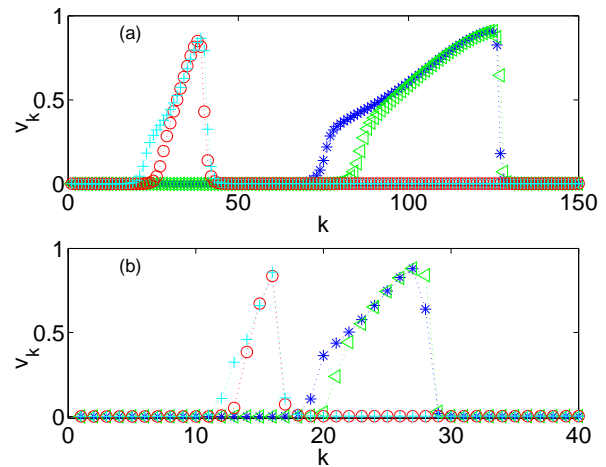}
\caption{Comparison of pulses for the  HH model
(\ref{hh1})-(\ref{hh5})  (circles),
HH with $h_k+n_k=h^*+n^*$ (crosses),
HH with $m_k=m_{\infty}(v_k)$ (triangles)
and the FHN-like system (\ref{fs1})-(\ref{fs2}) (asterisks):
(a) $D=0.09$, (b) $D=0.01$. The pulses have been generated
by a short stimulation at $k=0$ and have been superimposed
at a fixed time $t$. The pulse advances faster when
only one fast variable is kept (asterisks and triangles).
All variables are expressed in dimensionless units.}
\label{figura5}
\end{center}
\end{figure}

Numerical simulations show that reduced models involving only one
fast and one slow variables are quantitatively inaccurate.
Fig. \ref{figura5} compares pulse solutions of the full discrete
Hodgkin-Huxley model (circles) and a FitzHugh-Nagumo type reduction
(asterisks) generated by exciting the left end of a fiber.
The pulse solutions of the HH model are slower and narrower than
the pulse solutions of the FHN-like reduction.
Discarding one of the two fast variables produces a pulse with
an increased speed, as illustrated by Figure \ref{figura5}.
To describe the propagation of the pulse leading edge  we
need to keep the two fast variables.
Their evolution is described by a discrete bistable equation
coupled to an ordinary differential equation.
This system yields a Nagumo type equation (a single discrete
bistable equation) only in a very particular limit. Similarly,
discarding one of the  slow variables produces a slightly wider
pulse than the one described by the two original slow variables.
Moreover, FHN pulses and HH pulses have different structures.
Pulse solutions of FHN-like models typically consist of two
rigidly moving wave fronts \cite{cbfhn}.
HH pulses are ``triangular waves'': they are formed
by a leading wave front followed by a smooth  region, as shown in
Figure  \ref{figura5}.

In this paper we introduce an asymptotic strategy to construct
solitary pulses and wave trains in the discrete HH model. Our
asymptotic study exploits time scale separation to split the
variables in two blocks. The leading edge of the pulses is a wave
front solution of the reduced system involving the two fast
variables. This selects the speed of the wave. The two slow
variables become  relevant to determine the width of the peaks. Our
asymptotic constructions agree reasonably well with numerical
solutions for typical experimental data. For continuous HH models, a
quantitative approximation scheme exploiting time scale separation
was proposed in \cite{muratov}. In the discrete case, traveling
impulses  do not appear as solutions of ordinary differential
equations. Instead, more complicated differential-difference
equations have to be analyzed.

The paper is organized as follows.
In Section II we describe the discrete Hodgkin-Huxley model for myelinated
nerves. We present a few numerical solutions and discuss the reasons for
the poor performance of FitzHugh-Nagumo type reductions. Solitary pulses
are constructed in Section III. Section IV studies  wave train
generation at the boundary by periodic firing. In Section V, we describe
the two main mechanisms for propagation failure, related to damage in
the myelin sheath and the action of chemicals. Section VI  contains our
conclusions. In Appendix A we recall the derivation of the model.
Appendix B explains the nondimensionalization procedure.
Appendices C, D and E contain additional material on pulses.

\section{The Hodgkin-Huxley model for myelinated nerves}

\subsection{Dimensionless equations}

\label{sec:hh}

The dimensionless Hodgkin-Huxley (HH) model for a myelinated nerve
axon is:
\begin{eqnarray}
{dv_k \over dt} + I(v_k,m_k,n_k,h_k)
= D (v_{k+1}-2v_k+v_{k-1}), \label{hh1} \\
{dm_k \over dt} = \Lambda_m(v_k) \big[m_{\infty}(v_k)-m_k\big],
 \label{hh2} \\
{dn_k \over dt} = \epsilon \Lambda_n(v_k) \big[n_{\infty}(v_k)-n_k\big],
 \label{hh3} \\
{dh_k \over dt} = \epsilon \lambda \Lambda_h(v_k) \big[h_{\infty}(v_k)-
h_k\big],
 \label{hh4}
\end{eqnarray}
with
\begin{eqnarray}\begin{array}{lcl}
I(v,m,n,h)&=&g_K n^4 (v-{V}_K) +{g}_{Na} m^3
h(v-1) \\ &&+  {g}_{L} (v-{V}_L).
\end{array} \label{hh5}
\end{eqnarray}
Here, $v_k$ is the ratio of the deviation from rest of the membrane
potential to a reference potential, $n_k$ is the potassium activation,
$m_k$ is the sodium activation and $h_k$ the sodium inactivation.
Appendix A recalls the derivation of the model.
Appendix B details the procedure we have followed to nondimensionalize
the system.
The time scale is chosen by looking at the relaxation times $\tau_n$,
$\tau_m$ and $\tau_h$ for $n_k$, $m_k$ and $h_k$. Typically, $\tau_m
\ll \tau_n$ and $\tau_n\sim \tau_h$. Thus,  $\epsilon= {\tau_m\over
\tau_n}$ is small.

\begin{figure}
\begin{center}
\includegraphics[width=8cm]{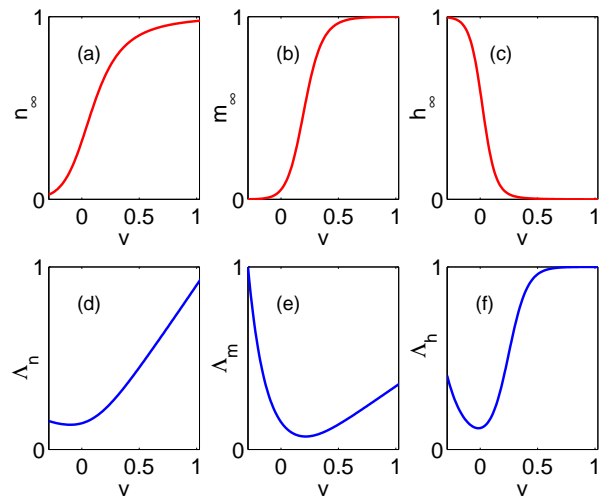}
\caption{Coefficient functions in the HH equations as functions
of $v_k=v$, in dimensionless units.}
\label{figura0}
\end{center}
\end{figure}

The rate functions and the stationary states in (\ref{hh2})-(\ref{hh4})
can be fitted to experimental data. Figure \ref{figura0}
plots their shape for the motor axon of a frog.
Analytic expressions for these functions and typical values of the
parameters for the frog nerve are collected in Appendix B and
will be used  in our numerical tests. The values for
$g_{Na},g_K,g_L$ have orders of magnitude $1, 10^{-1}, 10^{-2}$,
respectively.
The coupling parameter $D\sim 10^{-1}$ and $\epsilon \sim 10^{-2}$
with $\lambda \sim 1$. This means that we have two separate time
scales and that discreteness effects are relevant.

\subsection{Numerical solutions}
\label{sec:numerical}

\begin{figure}
\begin{center}
\includegraphics[width=8cm]{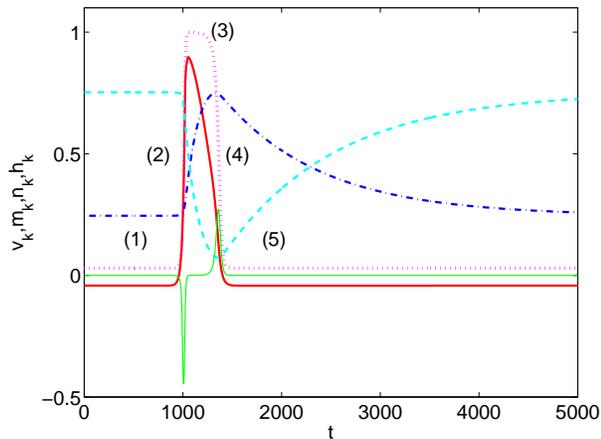}
\caption{Temporal action potential $v_k(t)$ (thick solid line)
generated by a single stimulation at $k=0$. The temporal
profiles of $m_k$ (dotted), $n_k$ (dashed-dotted), $h_k$ (dashed)
have been superimposed, all of them at a fixed node $k$.
The thin solid line shows the profile of $m_k-m_{\infty}(v_k)$.
All variables are expressed in dimensionless units.}
\label{figura2}
\end{center}
\end{figure}

\begin{figure}
\begin{center}
\includegraphics[width=8cm]{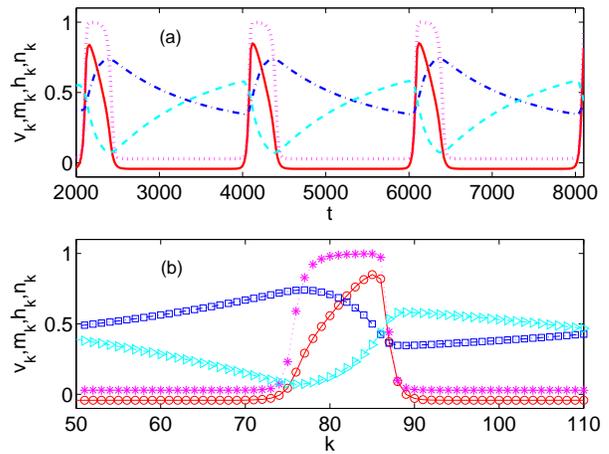}
\caption{(a) Same as in Fig. \ref{figura2} but generated by a
periodic stimulation at $k=0$. (b) Corresponding spatial profiles at
a fixed time: $v_k$ (circles), $m_k$ (asterisks), $n_k$ (squares),
$h_k$ (triangles). Zoom in one spatial period. All variables are
expressed in dimensionless units.} \label{figura6}
\end{center}
\end{figure}

System (\ref{hh1})-(\ref{hh4}) displays excitable behavior when it
has a unique constant stationary state $(v^*,m^*,n^*,h^*)$,
which is stable. %=(-0.04,0.029,0.24,0.75)
Figures \ref{figura2} and \ref{figura6} show solitary pulses
and wave train solutions generated by solving (\ref{hh1})-(\ref{hh4})
numerically for the parameter values in Appendix B.
After a short transient, the system relaxes to a traveling wave:
$v_k(t)=v(k-ct)$,  $m_k(t)=m(k-ct)$,  $n_k(t)=n(k-ct)$
and $h_k(t)=h(k-ct)$. All nodes undergo the same evolution
with a time delay ${1\over c}$.

As shown in Figures \ref{figura2} and  \ref{figura6}, pulses and wave
trains  are composed of  ``sharp'' interfaces and  ``smooth'' regions.
Let us describe the temporal profiles of a pulse as we move  from
left to right starting from the equilibrium region (1) in Figure \ref{figura2}.
The leading edge of each pulse is a sharp interface, marked as (2) in
Fig. \ref{figura2}. At this leading interface, $n_k$ and $h_k$ remain almost
constant whereas  $v_k$ and $m_k$ undergo abrupt changes. The
evolution of $v_k$ and $m_k$ is described by a reduced bistable system
and the leading edge is a wave front solution of this system.
A smooth region follows, where $n_k, h_k$ vary slowly and $v_k$ and $m_k$
are quasi-static: $m_k=m_{\infty}(v_k)$ and $I(v_k,m_{\infty}(v_k);n_k,m_k)=0$.
It is marked as (3) in Fig. \ref{figura2}.
At the end of this region the approximation $m_k=m_{\infty}(v_k)$
breaks down, see the thin solid line in Figure \ref{figura2}.
A trailing interface  develops, where $m_k$ changes abruptly  whereas
$n_k,h_k$ remain almost constant. This new region is marked as (4)
in Fig. \ref{figura2}. Notice that the fast variable $v_k$ happens to be
near its equilibrium value $v^*$ and decreases smoothly towards it.
Diffusion is negligible and
the evolution of $v_k$ and $m_k$ at the trailing edge is
governed by a system of ordinary differential equations. The bistable
structure is lost and no trailing wave front is formed.

The variables seem to be split in two blocks: slow
(evolving in the dimensionless time scale $T=\epsilon t$) and fast
(evolving in the dimensionless time scale $t$).
This suggest the possibility of finding an asymptotic description
of  pulses by exploiting time scale separation. The difficulty
of dealing with a discrete space variable is overcome by noticing
that the traveling wave profiles are smooth functions of a continuous
variable and solve a set of differential-difference equations
(see Appendix C).
Let us first check whether the number of variables can be reduced.

\subsection{Failure of simple FitzHugh-Nagumo type reductions}
\label{sec:compare}

It is quite tempting to look for a  simple  asymptotic description of
pulses in terms of one fast and one slow variable, as in the
FitzHugh-Nagumo model \cite{fhn,cbfhn}. These reduced models
usually assume that $m_k$ relaxes instantaneously to its equilibrium
state $m_k=m_{\infty}(v_k)$ and that the sum of the two slow variables
is  constant during an action potential $n_k+h_k=r$ \cite{keener2}:
\begin{eqnarray}
{dv_k \over dt}+ g_K n_k^4 (v_k\!-\!{V}_K) \!+
\!{g}_{Na} m_{\infty}(v_k)^3 (r-n_k)(v_k\!-\!1) \nonumber \\
\!+\! {g}_{L} (v_k\!-\!{V}_L) = D (v_{k+1}-2v_k+v_{k-1}), \label{fs1} \\
{dn_k \over dt} = \epsilon \Lambda_n(v_k)
\big[n_{\infty}(v_k)-n_k\big].
 \label{fs2}
\end{eqnarray}
In view of the numerical results described in Section \ref{sec:numerical},
these assumptions are inaccurate. The solid line in Fig. \ref{figura2}
shows that the difference $m_k-m_{\infty}(v_k)$  is not negligible at the
leading wave front. Setting $m_k=m_{\infty}(v_k)$ distorts the speed,
as shown in Figure \ref{figura5}.
On the other hand, $n_k+h_k$ is not constant during at action potential.
We may select $r=n^*+h^*$ to fit the leading edge but a slightly different
value of $r$ is required for the trailing part of the pulse.
Keeping $n_k+h_k=r$ during the action potential alters the width of the
pulse, as shown in Figure \ref{figura5}.

In conclusion, the FHN type reduction (\ref{fs1})-(\ref{fs2}) may be
useful to gain insight on pulse motion and propagation failure in excitable
media, but it is quantitatively inaccurate for the HH system with realistic
parameter values.

\section{Asymptotic construction of pulses}

\label{sec:pulses}

Accurate  descriptions of pulse waves in the HH model have to
deal with the full set of equations.  In this Section, we find
an approximation of their temporal profiles by matched
asymptotic expansions as $\epsilon \rightarrow 0$. This construction
yields predictions for the speed and width of the pulses, as well
as a characterization of the parameter ranges in which propagation fails.
We explain the procedure for the parameter values indicated in
(\ref{frog}). The structure of the pulse may change slightly for other
parameter values. Other possible structures are discussed in
Appendix \ref{sec:twointerfaces}.

In a traveling pulse we distinguish five regions (illustrated in Figure
\ref{figura2}). In each of them, a reduced description holds. The
whole profile is reconstructed by matching the approximated solutions
found in each region at zeroth order (see \cite{lager} for a description
of this technique). Our reference time scale is the slow time scale
$T=\epsilon t$. The technical details of the matching are given in Appendix
\ref{sec:matching}.

\begin{figure}
\begin{center}
\includegraphics[width=8cm]{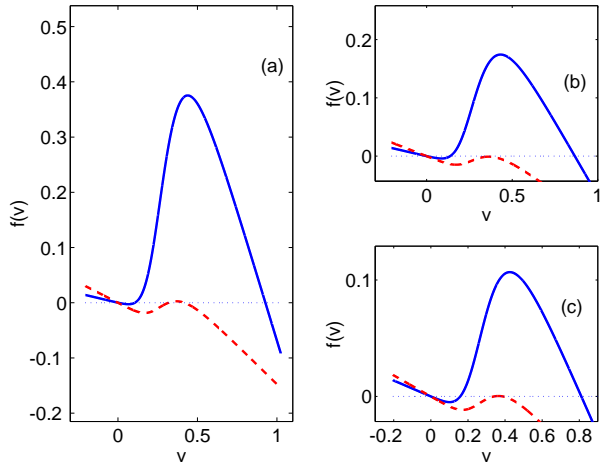}
\caption{Shape of the source $f(v)$: (a) $g_{Na}=1.49$, (b)
$g_{Na}=0.745$, (c) $g_{Na}=0.497$, when $(n,h)=(n^*,h^*)$ (solid
line) and $(n,h)=(n^{[1]},h^{[1]})$ (dashed line). All variables are
expressed in dimensionless units.} \label{figura1}
\end{center}
\end{figure}

The resulting temporal pulse profiles with $k$ fixed have the following
structure:

\begin{enumerate}

\item {\it Front of the pulse}. This is region (1) in Figure
\ref{figura2}. Here  $v_k\sim v^*$,
$m_k\sim m^*$, $n_k\sim n^*$ and $h_k\sim h^*$,
$(v^*,m^*,n^*,h^*)$ being the stable equilibrium
state of the system.

\item {\it Leading edge of the pulse}, located at $T=T_0$ and
marked as (2) in Figure \ref{figura2}.
The slow variables, $n_k$ and $h_k$, remain essentially constant but the
fast variables, $v_k$ and $m_k$, undergo rapid changes in the time scale
$\overline{t}={T-T_0\over \epsilon}\in(-\infty,\infty)$.
To leading order,
\begin{eqnarray}\begin{array}{lll}
{dv_k \over d\overline{t}} + I(v_k,m_k,n_k,h_k) = D (v_{k+1}-2v_k+v_{k-1}), \\
{dm_k \over d\overline{t}} = \Lambda_m(v_k) \big[m_{\infty}(v_k)-m_k\big], \\
{dn_k \over d\overline{t}} = 0, \quad
{dh_k \over d\overline{t}} = 0,
\end{array}\label{tw1} \end{eqnarray}
Thus, $n_k$ and $n_k$ remain almost constant. Matching with the
previous region, $n_k=n^*$ and $h_k=h^*$ (see Appendix \ref{sec:matching}).
The evolution of $v_k$ and $m_k$ is described by the `fast reduced
system':
\begin{eqnarray}
{dv_k \over d\overline{t}} + I(v_k,m_k,n_k,h_k)
= D (v_{k+1}-2v_k+v_{k-1}), \label{f1} \\
{dm_k \over d\overline{t}} = \Lambda_m(v_k) \big[m_{\infty}(v_k)-m_k\big],
\label{f2}
\end{eqnarray}
with  $n_k=n^*$ and $h_k=h^*$. This system  displays bistable behavior.
Let us denote by $v=\nu^{(i)}(n,h), i=1,2,3$ the three solutions of
\begin{eqnarray}
f(v;n,h)= -I(v, m_{\infty}(v),n,h)=0
\label{r1}
\end{eqnarray}
in a neighborhood of $(n^*,h^*)$. The system (\ref{f1})-(\ref{f2})
has a unique stable traveling wave front solution
$v_k(t)=v(k-c\overline{t})$,
$m_k=m(k-c\overline{t})$   joining the two stable constant
states $\nu^{(3)}(n^*,h^*)$, $m_{\infty}(\nu^{(3)})$  and
$\nu^{(1)}(n^*,h^*)$, $m_{\infty}(\nu^{(1)})$, which propagates
with a definite speed  $c=c_+(n^*,h^*)$.
This wave front solution is the leading edge of the pulse.

\item {\it Top of the pulse}. This is region (3) in Figure
\ref{figura2}. Here, the slow variables evolve in the time scale $T$ and
the fast variables relax instantaneously to their equilibrium values:
$m_k=m_{\infty}(v_k)$, in which $v_k$ solves $f(v_k;n_k,h_k)=0$.
The evolution of the slow variables is governed by the 'slow reduced
system':
\begin{eqnarray}\begin{array}{l}
{dn_k \over dT} \!=\!  \Lambda_n(v_k)
\big[n_{\infty}(v_k) \!-\! n_k \big], \\
{dh_k \over dT} \!=\!  \lambda \Lambda_h(v_k)
\big[h_{\infty}(v_k)  \!-\! h_k \big],
\end{array}\label{s1}
\end{eqnarray}
for $T_0<T<T_1$.
Matching with the previous stage we get $v_k= \nu^{(3)}(n_k,h_k)$,
$n_{k}= n^*$ and $h_{k}= h^*$ at $T=T_0$ (see Appendix
\ref{sec:matching}).
For the parameter values indicated in (\ref{frog}), the third
branch of roots $v_k= \nu^{(3)}(n_k,h_k)$ of  $f(v_k;n_k,h_k)=0$ disappears
at $(n_k,h_k)=(n^{[1]},h^{[1]})$, colliding with the second branch:
 $\nu^{(3)}(n^{[1]},h^{[1]})=\nu^{(2)}(n^{[1]},h^{[1]})=v^{[1]}$.
At the corresponding time $T=T_1$, region
(3) ends. The time $T_1$ is characterized by $n_k(T_1)=n^{[1]}$
and $h_k(T_1)=h^{[1]}$ \cite{nota}.

\item {\it Trailing edge of the pulse}, located at $T_1$ and marked
as (4) in Figure \ref{figura2}.
In this region, $v_k$ is no longer at equilibrium since the
two largest roots of $f(v;n_k,h_k)=0$ are lost.
The fast variables evolve in the time scale $\overline{t}={T-T_1\over \epsilon}
\in(-\infty,\infty)$, whereas the slow variables remain essentially constant.
Matching with the previous stage, $n_k=n^{[1]}$, $h_k=h^{[1]}$.
Due to the particular  shape of $f$ (see the dashed line in
Figure \ref{figura1} (a)), $\nu^{(2)}(n^{[1]},h^{[1]})$ is close enough
to $\nu^{(1)}(n^{[1]},n^{[1]})$ for $v_k-v_{k-1}$ to be small.
Thus, we may neglect the discrete differences $D(v_{k+1}-2v_k+v_{k-1})$ and
the fast variables are governed by a system of ordinary differential
equations:
\begin{eqnarray}\begin{array}{l}
{dv_k \over d\overline{t}}= - I(v_k,m_k,n^{[1]},h^{[1]}),   \\
{dm_k \over d\overline{t}}= \Lambda_m(v_k) \big[m_{\infty}(v_k)-m_k\big].
\end{array} \label{edo}
\end{eqnarray}
This system has one stable equilibrium point: $
v^{[2]}=\nu^{(1)}(n^{[1]},h^{[1]}),$ $m^{[2]}=m_{\infty}(v^{[2]})$.
The fast variables evolve from their initial values $v_k=v^{[1]}$
and $m_k=m^{[1]}=m_{\infty}(v^{[1]})$ to the equilibrium point.

\item {\it Pulse tail}. This is region (5) in Figure \ref{figura2}.
In the pulse tail, the fast variables relax instantaneously to
their equilibrium values: $m_k=m_{\infty}(v_k)$  and
$v_k=\nu^{(1)}(n_k,h_k).$ The slow variables solve (\ref{s1}) with
$v_k=\nu^{(1)}(n_k,h_k)$ for $T_1< T <\infty$ and evolve smoothly from
$(n^{[1]},h^{[1]})$ to their equilibrium values $(h^*,n^*)$ as
$T\rightarrow \infty$.

\end{enumerate}

Uniform approximations to the temporal profiles obtained by gluing
together the approximated solutions in each region are given in
Appendix \ref{sec:matching}. Figure \ref{figura9} compares the
asymptotic reconstruction to the actual profiles. The agreement
improves as $\epsilon$  decreases.

\begin{figure}
\begin{center}
\includegraphics[width=8cm]{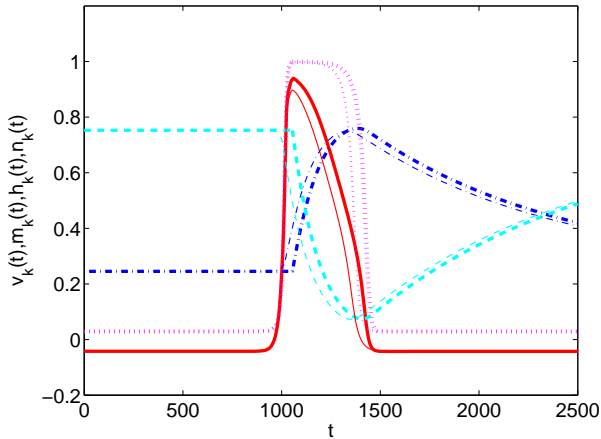}
\caption{Asymptotic reconstruction of the temporal profiles
of $v_k(t)$ (thick solid line), $m_k$ (thick dotted line), $n_k$
(thick dashed-dotted line), $h_k$ (thick dashed line),
versus the actual temporal profiles (thin lines).
All variables are expressed in dimensionless units.}
\label{figura9}
\end{center}
\end{figure}

Our  construction characterizes the speed of the traveling pulse: it
is the speed of the wave front solution of (\ref{f1})-(\ref{f2})
with $n_k\sim n^*$ and $h_k\sim h^*$. The speed of these fronts can
be predicted by a depinning analysis for small speeds (as in
\cite{cbn}) or in terms of continuous waves for large speeds (as in
\cite{keener1}).

Let us now compute the width of the pulse. In the peak, $(n_k,h_k)$ lies
in the integral curve of:
\begin{eqnarray}\begin{array}{l}
{d h \over dn} =
{\lambda \Lambda_h \big(\nu^{(3)}(n,h)\big)
\big[h_{\infty}\big(\nu^{(3)}(n,h)\big)-h \big] \over
  \Lambda_n\big(\nu^{(3)}(n,h)\big)
\big[n_{\infty}\big(\nu^{(3)}(n,h)\big)-n \big]},
\end{array}\label{w1}
\end{eqnarray}
selected by $h(n^*)=h^*$ and $h(n^{[1]})=n^{[1]}$.
Going back to the time
scale $t={T\over \epsilon}$ in (\ref{s1}), the peak width is:
\begin{eqnarray}\begin{array}{l}
 {\cal T}_1= \!{T_1-T_0\over \epsilon}=\!\int_{n^*}^{n^{[1]}} \!\!
{ds \over \epsilon \Lambda_n\big(\nu^{(3)}(s,h(s))\big)
\big[n_{\infty}\big(\nu^{(3)}(s,h(s))\big)-s \big]}
\end{array}\label{w2bis}
\end{eqnarray}
where the integral is calculated along the solution $h(n)$ of (\ref{w1}).
Similarly, the duration of tail is:
\begin{eqnarray}
{\cal T}_2= \int_{n^{[1]}}^{n^*} {ds \over \epsilon
\Lambda_n\big(\nu^{(1)}(s,h(s))\big)
\big[n_{\infty}\big(\nu^{(1)}(s,h(s))\big)-s \big]}, \label{w5bis}
\end{eqnarray}
with
\begin{eqnarray}\begin{array}{l}
{d h \over dn} =
{\lambda_h \Lambda_h \big(\nu^{(1)}(n,h)\big)
\big[h_{\infty}\big(\nu^{(1)}(n,h)\big)-h \big] \over
  \Lambda_n\big(\nu^{(1)}(n,h)\big)
\big[n_{\infty}\big(\nu^{(1)}(n,h)\big)-n \big]}, \; h(n^{[1]})=h^{[1]}.
\end{array}\nonumber
\end{eqnarray}
The integral ${\cal T}_2$ diverges due to a singularity at $n^*$.
However, we can use it to predict how long does it take for
the tail to get close enough to $n^*,h^*$  replacing $n^*$ by
$n^*-\eta$.

The spatial length of the peak is found using the traveling wave
structure: $v_k(t)=v(k-ct).$ Thus, ${\cal T}_1$ is the time elapsed
from the instant at which the leading front reaches a point $k$ to
the instant when the end of the peak crosses the same point $k$. The
number of nodes $L_1$ in the pulse peak is approximately the integer
part of:
\begin{eqnarray}\begin{array}{l}
L_1\sim  c_+(n^*,h^*) {\cal T}_1.
\end{array}\label{w2}
\end{eqnarray}
Our asymptotic construction is consistent when  $L_1\geq 1$. This
yields a restriction on the size of $\epsilon$ for the existence
of pulses:
\begin{eqnarray}
\epsilon \leq  c_+ \!  \int_{n^*}^{n^{[1]}}\!\!
{ds \over  \Lambda_n\big(\nu^{(3)}(s,h(s))\big)
\big[n_{\infty}\big(\nu^{(3)}(s,h(s))\big)-s \big]}.
\label{w3}
\end{eqnarray}
A similar argument can be applied in the infinite tail to
quantify the number of nodes at which $n_k,h_k$ depart
noticeably from equilibrium.

The values predicted by our asymptotic theory for the parameters in
Appendix B are $c=c_+=0.069$, $n^{[1]}=0.777$, $h^{[1]}=0.099$,
$L_1=12$, in good  agreement with the numerical measurements.
The leading and trailing edges contain about $4$ more nodes,
 that we have neglected.

\begin{figure}
\begin{center}
\includegraphics[width=8cm]{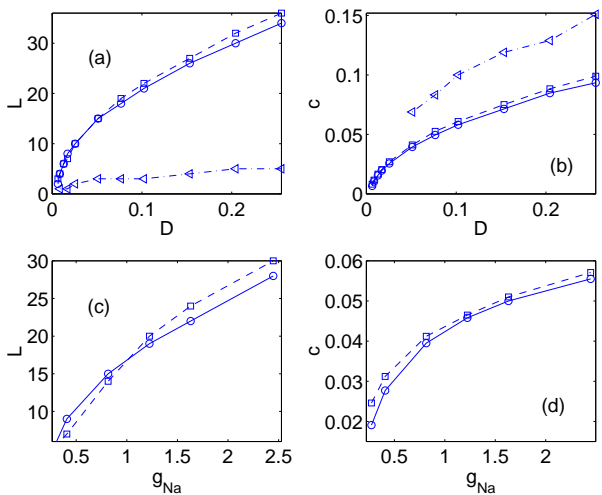}
\caption{Spatial width of the peak (circles) compared to the predicted
width (squares) as a function of: (a) $D$ and (c) $g_{Na}$.
In (a), triangles represent the number of additional nodes forming the
leading edge;
Speed of the pulse (circles) compared to the predicted speed (squares)
as a function of: (b) $D$ and (d) $g_{Na}$. In (b), triangles represent
the speed of the FHN  reduction.
All variables are expressed in dimensionless units.}
\label{figura8}
\end{center}
\end{figure}

Figure \ref{figura8} compares the predicted and numerically measured
widths and speeds as the parameters $D$ and $g_{Na}$ change.
Notice that the values $(n^*,h^*)$ and $(n^{[1]},h^{[1]})$ are independent
of $D$. Thus, the dependence of the width on $D$ comes through the
speed. This explains the similar curves observed in Figure \ref{figura8}
(a) and (b). The triangles in Figure \ref{figura8} (a) represent the
number of nodes in the leading edge, that must be added to compute
the total length of a peak.
Changes in $g_{Na}$ affect the curves $\nu^{(i)}(n,h)$, $i=1,2,3$, and
the speed of the fronts in the reduced fast system. As Figures
\ref{figura8} (c) and (d) show, the quantitative agreement between
predicted and numerically measured widths and speeds is reasonable,
as long as we are not close to the critical thresholds $D_c$ and $\epsilon_c$
for propagation failure. We will discuss this point further in Section
\ref{sec:failure}.

For the parameters indicated in (\ref{frog}), FHN type reductions
perform poorly. The widths predicted as $D$ changes are particularly bad.
We find 160 or 250 points in the peak, when 18 or 34 are expected.
The predictions for the speed are better, see the triangles in Figure
\ref{figura8} (b).
Similar comments apply when $g_{Na}$ is modified.

\section{Asymptotic construction  of wave trains}

\label{sec:wavetrains}

When a nerve fiber is periodically excited, we expect propagation
of signals in form of wave trains.
Wave trains resemble a sequence of identical pulses periodically
spaced \cite{keener3}. The asymptotic description of each of these pulses
is similar to the construction of solitary pulses given in Section
\ref{sec:pulses}. However, there are several differences. First, the front
of the pulses is another pulse and not a region at equilibrium.
Second, the leading edge is a wave front solution of the fast
reduced system (\ref{f1})-(\ref{f2}) with $n_k=N$, $h_k=H$,
$(N,H)\neq (n^*,h^*)$. This wave front solution selects the speed
$c=c_+(N,H)$ of the wave train. It has to match the tail of the previous
pulse, described by the slow reduced system (\ref{s1}) with
$v_k=\nu^{(1)}(n_k,h_k)$. Thus, we find a  family of wave trains for
couples $(N,H)$ lying on an integral curve of:
\begin{eqnarray}
{d h \over dn} = {\lambda \Lambda_h(\nu^{(1)}(n,h)) \big[h_{\infty}
(\nu^{(1)}(n,h))-h\big] \over  \Lambda_n(\nu^{(1)}(n,h))
\big[n_{\infty}(\nu^{(1)}(n,h))-n\big]}.
\label{wt1}
\end{eqnarray}
The curve is selected by observing that each pulse in the wave train
has one driving leading edge, as in Section \ref{sec:pulses}.
This means that $h(n^{[1]})= h^{[1]}$, where $(n^{[1]},h^{[1]})$
are the values for which $\nu^{(2)}(n^{[1]},h^{[1]})=\nu^{(3)}(n^{[1]},h^{[1]})$.

The spatial period is approximated by adding the lengths of the smooth regions:
$L(N,H)= c_+(N,H) {\cal T(N,H)}$ where
${\cal T}(N,H)={\cal T}_1(N,H)+{\cal T}_2(N,H)$ is the
time spent in the excited and recovery branches. The temporal lengths
${\cal T}_1, {\cal T}_2$ are
defined as in Section \ref{sec:pulses}, with $n^*,h^*$ replaced by $N,H$.
Now, ${\cal T}_2(N,H)$ is finite.

The values predicted by our asymptotic theory for Figure \ref{figura6}
are $c=c_+=0.052$, $n^{[1]}=0.755$, $h^{[1]}=0.09$, $L\sim 37-38$,
in good agreement with the numerical measurements.
Numerically, we obtain $L\sim 43$ for the spatial period. The difference is
approximately the number of points in the  edges, that we have neglected.

\section{Propagation failure}

\label{sec:failure}

\begin{figure}
\begin{center}
\includegraphics[width=8cm]{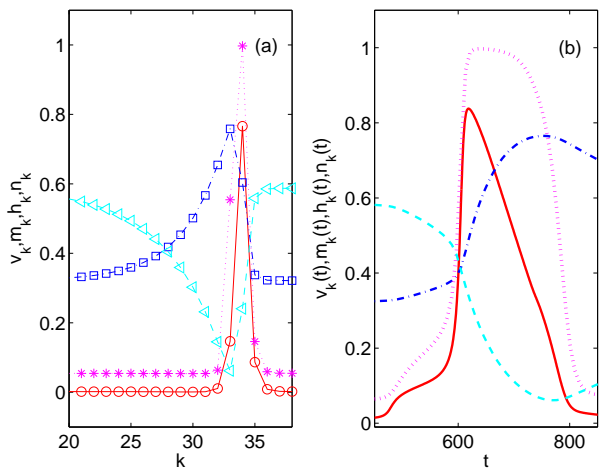}
\caption{Action potential  near propagation failure:
(a) Spatial profiles of $v_k$ (circles), $m_k$ (asterisks),
$n_k$ (squares) and $h_k$ (triangles) superimposed at $t$ fixed,
(b) Temporal profiles of $v_k(t)$ (solid), $m_k(t)$ (dotted), $n_k(t)$
(dotted-dashed) and $h_k(t)$ (dashed) superimposed for $k$ fixed.
All variables are expressed in dimensionless units.}
\label{figura4}
\end{center}
\end{figure}

Pulses may fail to propagate at least by two reasons: weak coupling
and inadequate time scale separation between the different variables.
Weak couplings cause propagation failure by pinning the leading
edge of the pulse. A small time scale separation between the two
blocks of slow and fast variables produces pulses of vanishing
width.
Other scenarios for failure might arise when more than
two time scales are present. We focus here on mechanisms for failure
associated with changes in the parameters due to illness or drugs.

\subsection{Propagation failure due to damage in the myelin sheath}

\label{sec:myelin}

The loss of myelin alters the value of the coupling coefficient $D$.
Then, the leading wave front can only propagate if the  $D$
is large enough to avoid the pinning in the fast reduced system
(\ref{f1})-(\ref{f2}) with $n_k=n^*$ and $h_k=h^*$. The
critical coupling $D_c$ depends on the shape of the nonlinear sources,
which is controlled by the parameters $g_{Na}$,$g_{K}$,$g_{L}$,
$V_{Na}$,$V_{K}$ and $V_{L}$. The general rule is that
the area  $A_{23}$ enclosed by $f(v)$ between its second and the
third zeroes has to be large enough (depending on $D$) for the leading edge
to propagate. The proximity of failure is detected by the fact that the
wave profiles develop `steps'.  Figure \ref{figura4}(b) illustrates
generation of `steps' in the time profile of the pulse for $v_k$ near
propagation failure  $D=0.0072 \sim D_c$.  As for bistable equations,
the depinning transition is associated with a bifurcation in the
system \cite{cbn}.

\subsection{Propagation failure due to the action of chemicals}

\label{sec:chemicals}

\begin{figure}
\begin{center}
\includegraphics[width=8cm]{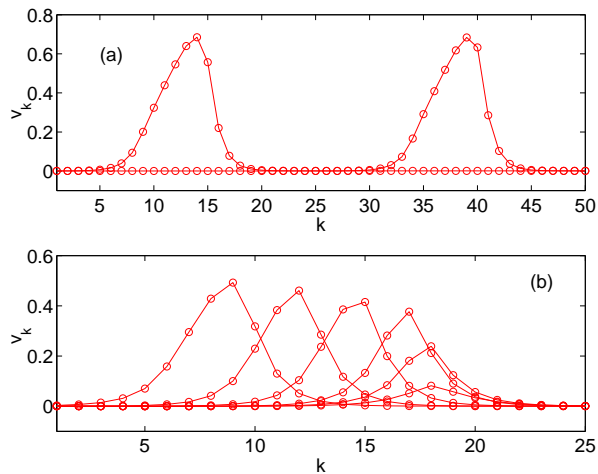}
\caption{Impact of decreasing $g_{Na}$ on pulse propagation: (a)
$g_{Na}=0.745$, (b) $g_{Na}=0.497$. The spatial action potentials
have been superimposed at different times. All variables are
expressed in dimensionless units.} \label{figura3}
\end{center}
\end{figure}

Many drugs block the propagation of nerve impulses by reducing sodium and
potassium conductivities \cite{scott1,keener2}.
Figure \ref{figura1}(b) illustrates the impact of decreasing $g_{Na}$
on the shape of $f$. The enclosed area decreases and so does the propagation
speed, up to a critical value at which the width of the pulses (given by
formula (\ref{w2})) vanishes.
Only decremental pulses are observed, see Figure \ref{figura3}.

\section{Conclusion}

\label{sec:conclusion}

We have introduced an asymptotic strategy to construct solitary
pulses and wave trains in the discrete Hodgkin-Huxley model.
Unlike FHN reductions, our asymptotic descriptions agree
reasonably well with numerical solutions of the discrete HH system.
We have discussed two mechanisms for propagation failure.
The first one is related to damage in the myelin sheath and is
reminiscent of the depinning transitions for wave fronts in
bistable systems \cite{cbn} as the coupling decreases.
The second mechanism reflects the impact of chemicals
on the time scale separation in the model.
Our analysis applies to isolated nerve fibers. However, real motor
nerves of vertebrates  comprise several hundred of interacting
fibers \cite{scott2}. It would be interesting to extend our
asymptotic predictions to bundles of fibers.

Our asymptotic construction may be useful to understand systems with
a mathematical structure similar to HH: models for propagation
of impulses through cardiac tissue \cite{beeler}, models of charge
transport in semiconductor superlattices \cite{bonilla} or the more
complex Frankenhauser-Huxley model for myelinated nerves \cite{goldman}.

\acknowledgments

This work has been supported by the Spanish MCyT through grant
BFM2002-04127-C02-02, and by the European Union under grant
HPRN-CT-2002-00282. The author thanks one of the unknown referees
for bringing reference \cite{muratov} to her attention.

%Problematic units: $F=As/V$, $S=A/V=mho$
%$(s^{-1}\mu F /cm^2)/ (mmho/cm^2)= 10^{-6} F/(s mmho)=
%10^{-3} F/(s A/V)=10^{-3}$
%$(\mu A/cm^2) / (mV mmho/cm^2) =  A / (V A / V)=1 $

\appendix

\section{The discrete Hodgkin-Huxley model for myelinated nerves}
\label{sec:model}

\begin{figure}
\begin{center}
\includegraphics[width=8cm]{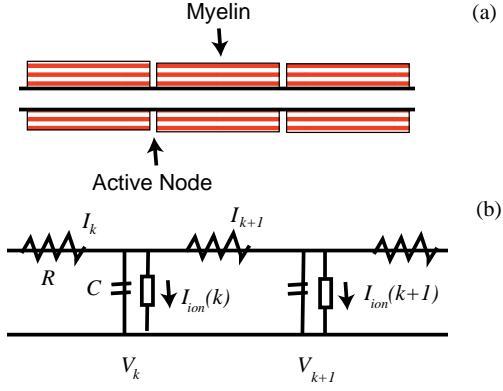}
\caption{(a) Structure of a myelinated nerve fiber (not to scale).
(b) Equivalent circuit for a myelinated nerve fiber.}
\label{figura7}
\end{center}
\end{figure}

Myelinated nerve fibers are covered almost everywhere by an insulating
coat of myelin. Only at small active sites (nodes of Ranvier) can
the membrane function in the normal way. Figure \ref{figura7}(a)
illustrates the structure of a myelinated fiber. The length of the
myelin sheath is typically $1$ to $2$ $mm$ (close to $100d$ where $d$
is the fiber diameter) and the width of the nodes is about $1 \mu m$.
The nodes have conduction properties similar to the unmyelinated nerve
membrane, while the myelin has a much higher resistance and lower
capacitance than the axonal membrane \cite{keener2}.
Myelinated nerve fibers can be described by a linear diffusion equation
which is periodically loaded by the active nodes
\cite{pickard66,markin1967,fh}. This picture can be simplified
by lumping the internode capacitance of the myelin together with the
nodal capacitance \cite{scott1}.
The myelin is considered to be a perfect insulator.
This leads to the equivalent circuit in
Figure \ref{figura7}(b). $C$ and $R$ represent lumped resistance and
capacitance. $V_k$, $I_k$ and $I_{ion}(k)$ represent the membrane potential,
internodal current and ionic current at the $k$-th node. Applying
Kirchoff's laws to the circuit yields:
\begin{eqnarray}
V_{k-1}-V_k= R I_k, \quad
I_k-I_{k+1}=C {d V_k \over dt} +I_{ion}(k) \label{kirchoff}
\end{eqnarray}
Adopting at each node the Hodgkin-Huxley expression for the ion current
\cite{hh},
we obtain the discrete Hodgkin-Huxley model for a myelinated axon:
\begin{eqnarray}\begin{array}{l}
C  {d V_k\over dT} + I_{ion}(V_k,M_k,N_k,H_k) =    \\
\overline{D} (V_{k+1}-2V_k+V_{k-1}),
\end{array}\label{apa0}\\
\begin{array}{l}
{d M_k \over dT} =\overline{\lambda}_M   \overline{\Lambda}_M(V_k)
(M_{\infty}(V_k)-M_k),    \\
{d N_k \over dT} =\overline{\lambda}_N   \overline{\Lambda}_N(V_k)
(N_{\infty}(V_k)-N_k),   \\
{d H_k \over dT} =\overline{\lambda}_H   \overline{\Lambda}_H(V_k)
(H_{\infty}(V_k)-H_k),
\end{array}\label{apa1} \end{eqnarray}
where the index $k$ designs the $k$-th node of the fiber.
Here, $V_k$ is the deviation from rest of the membrane potential,
$N_k$ is the potassium activation, $M_k$ is the sodium activation and
$H_k$ the sodium inactivation. The ion current is given by:
\begin{eqnarray}\begin{array}{l}
I_{ion}(V,M,N,H)= \overline{g}_{Na} M^3 H (V-\overline{V}_{Na,R})
\\+\overline{g}_L(V-\overline{V}_{L,R})
+ \overline{g}_K N^4 (V-\overline{V}_{K,R}). \\
\end{array}\label{apa2} \end{eqnarray}
The fraction of open $K^+$ channels is computed as $N_k^4$. The fraction
of open $Na^{+}$ channels is approximated by $M_k^3 H_k$.
The parameters have the following interpretation.
$\overline{g}_{Na}$ and $\overline{g}_K$ are the
maximum conductance values for $Na^{+}$  and $K^+$ pathways, respectively.
$\overline{g}_L$ is a constant leakage conductance. The corresponding
equilibrium potentials are $ \overline{V}_{Na}$, $\overline{V}_K$ and
$\overline{V}_L$, respectively.
Then, $\overline{V}_{Na,R}= \overline{V}_{Na}-\overline{V}_{R} $,
$\overline{V}_{K,R}=\overline{V}_{K}-\overline{V}_{R}$ and
$\overline{V}_{L,R}=\overline{V}_{L}-\overline{V}_{R},$
where $\overline{V}_R$ is the resting potential.
$C $ is the membrane capacitance. The coefficient
$\overline{D}={1\over L(r_i+r_e)}={1\over R}$, where $L$ is the length
of the myelin sheath between nodes and $r_i,r_e$ the resistances per unit
length of intracellular and extra-cellular media.

This model is adequate for the long axons of peripheral myelinated nerves.
More biological detail can be included by adding an equation for
the membrane potential $V(x,t)$ across the myelin sheath in the internodes
\cite{fh}:
\begin{eqnarray}
c {\partial V \over \partial T} =
{1\over r_i+r_e}{\partial^2 V \over \partial^2 x} - {V\over r}, \quad
x\in (x_k,x_{k+1}), t>0\label{fitz1}\\
V(x_k,t)=V_k(t), \; V(x_{k+1},t)=V_{k+1}(t) \label{fitz2}
\end{eqnarray}
coupled with (\ref{apa1}) and:
\begin{eqnarray}
C {dV_k \over dT} + I_{ion}(V_k,M_k,N_k,H_k)= I_k(t),
\label{fitz3} \\
I_k(t)= {1\over r_i+r_e}[{\partial V \over \partial x}(x_k^+,t)  -
{\partial V \over \partial x}(x_k^-,t) ].
\label{fitz4}
\end{eqnarray}
This model produces a good quantitative approximation of the conduction
velocity for toad axons \cite{fh}. Numerical simulations of
the sensitivity to different parameters (diameter, nodal area...) produce
results in agreement with experiments \cite{moore78,goldman}.
The discrete model (\ref{apa0})-(\ref{apa1}) is recovered by assuming that
the axial currents along the myelin sheath ${\partial V \over \partial x}
(x,t)$ are constant in each internode. Then, ${\partial V \over \partial x}
(x,t)={V_{k+1}(t)-V_{k}(t)\over L}$ in $[x_k,x_{k+1}]$ with $L=x_{k+1}-x_k$.
As a result, $I_k(t)={1\over L( r_i+r_e)}(V_{k+1}-V_k+V_{k-1})$.
This approximation is reasonable in view of the numerical results in
\cite{fh} (see Figure 2 therein).

\section{Dimensionless equations}
\label{sec:dimensionless}

For our numerical tests we have selected the parameters and coefficient
functions of a frog. For the motor nerve of a frog, the data in \cite{cole}
can be fitted by the following rate and stationary state functions:
\begin{eqnarray}
\begin{array}{l}
\overline{\Lambda}_M(V)\!=\! 0.03\big[{2.5 -0.1 V \over
\exp(2.5-0.1 V) -1} + 4 \; \exp({-V \over 18})\big],   \\
M_{\infty}(V)\!=\!\big[ 1+  4 \exp({-V \over 18})
{ \exp(2.5-0.1V) -1  \over (2.5 -0.1V)} \big]^{-1},\\
\overline{\Lambda}_H(V)\!=\!  \big[ 0.07 \exp({- V\over 20}) +
{1 \over \exp(3-0.1V ) +1}\big],   \\
H_{\infty}(V)\!=\!\big[ 1+  {\exp({V\over 20})
 \over 0.07 \big(\exp(3-0.1V) +1\big)   } \big]^{-1},\\
\overline{\Lambda}_N(V)\!=\! 0.79 \big[ {0.1-0.01V \over
\exp(1-0.1 V ) -1} + 0.125 \; \exp({-V \over 80})\big],  \\
N_{\infty}(V)\!=\!\big[ 1+   0.125  \exp({-V \over 80})
{ \exp(3-0.1 V) -1  \over (0.1 -0.01V)} \big]^{-1}.
\end{array} \label{apa3}
\end{eqnarray}
Typical values of the parameters  \cite{cole,scott1} are given below:
\begin{eqnarray}
\begin{array}{|c|c|c|c|}
\hline
\overline{D} & \overline{g}_{Na}& \overline{g}_K &
\overline{g}_{L} \\
\hline
 1/28  (M  \Omega)^{-1} & 0.57 \;  \mu mho &
0.104 \; \mu mho & 0.025 \; \mu mho \\
\hline
\overline{V}_R  & \overline{V}_{Na} & \overline{V}_K & \overline{V}_L \\
\hline
-75 \;mV &  47 \;mV & -75 \;mV  & -75 \;mV \\
\hline
C  & \overline{\lambda}_{M}  & \overline{\lambda}_{H} & \overline{\lambda}_{N} \\
\hline
2.6-4.7 \;p F & 127 \;(ms)^{-1} & 1.76 \;(ms)^{-1} & 2  \;(ms)^{-1} \\
\hline
\end{array} \nonumber
\end{eqnarray}
We nondimensionalize the model  by choosing as new variables
$v_k={V_k \over  \overline{V}_{Na,R} }$,
$t=T \overline{\lambda}_{M}$, $m_k=M_k$,
$n_k=N_k$, $h_k=H_k$. The dimensionless equations are:
\begin{eqnarray} \begin{array}{l}
{dv_k \over dt}+   g_K n_k^4 (v_k-{V}_K) +{g}_{Na} m_k^3
 h_k(v_k-1)+  \\ {g}_{L} (v_k-{V}_L)
= D (v_{k+1}-2v_k+v_{k-1}), \\
{dm_k \over dt} = \Lambda_m(v_k) \big[m_{\infty}(v_k)-m_k\big], \\
{dn_k \over dt} = \lambda_n \Lambda_n(v_k) \big[n_{\infty}(v_k)-n_k\big], \\
{dh_k \over dt} = \lambda_h \Lambda_h(v_k) \big[h_{\infty}(v_k)-h_k\big].
\end{array}\label{apa4}\end{eqnarray}
Set $G=C  \overline{\lambda}_{M}$. Then, the
dimensionless parameters are given by:
\begin{eqnarray}
\begin{array}{|c|c|c|c|c|c|c|c|c|}
\hline
{g}_{Na} & {g}_{K} & {g}_{L} & D & {V}_K & {V}_L &  \lambda_n
& {\lambda}_m  \\
\hline
{\overline{g}_{Na} \over G} & {\overline{g}_{K} \over G} &
{\overline{g}_{L} \over G} & {\overline{D} \over G} &
 {\overline{V}_{K,R} \over \overline{V}_{Na,R}} &
{\overline{V}_{L,R} \over \overline{V}_{Na,R}} &
{\overline{\lambda}_N \over \overline{\lambda}_{M}} &
{\overline{\lambda}_H \over \overline{\lambda}_{M}} \\
\hline
\end{array} \nonumber
\end{eqnarray}
The new rate functions and stationary states are obtained from (\ref{apa3})
replacing $V$ by $v \overline{V}_{Na,R}$.
In dimensionless units, the parameters for a frog nerve become:
\begin{eqnarray}
\begin{array}{|c|c|c|c|c|c|c|c|}
\hline
 {D} &  {g}_{Na}&  {g}_K & {g}_{L} & {V}_K &  {V}_L & {\lambda}_{h} &
{\lambda}_{n} \\ \hline
0.093 & 1.49  & 0.27 & 0.065  & 0 &  0 & 0.014  & 0.015   \\
\hline
\end{array} \label{frog}
\end{eqnarray}

In our asymptotic analysis, we choose $\lambda_n=\epsilon$ as small
parameter and write $\lambda_h=\epsilon \lambda$, $\lambda={\lambda_h \over
\lambda_n} \sim 1$.

\section{Equations for the wave profiles}
\label{sec:eigenvalue}

The wave profiles and speeds solve an eigenvalue problem for a
system of differential-difference equations:
\begin{eqnarray}\begin{array}{lll}
-c v_z(z) &=& D (v(z+1)- 2v(z)+v(z-1)) \\
& & -  I(v(z),m(z),n(z),h(z)), \\
-c m_z(z) &=&  \Lambda_m(v(z)) \big[m_{\infty}(v(z))-m(z)\big],\\
-c n_z(z) &=& \epsilon \Lambda_n(v(z)) \big[n_{\infty}(v(z))-n(z)\big],\\
-c h_z(z) &=& \epsilon \lambda \Lambda_h(v(z)) \big[h_{\infty}(v(z))-h(z)\big].
\end{array}\label{apa14}  \end{eqnarray}
In a solitary pulse, the profiles tend to the equilibrium states
as $z\rightarrow \pm \infty$. In a wave train, the profiles
are periodic: $v(z)=v(z+L)$, $m(z)=m(z+L)$, $n(z)=n(z+L)$ and
$h(z)=h(z+L)$, $L$ being the spatial period.

\section{Reconstruction of the temporal profile}

\label{sec:matching}

We describe below the matching conditions and the uniform
reconstruction of the pulse profiles in the different regions.
The superscripts (I),...,(V) refer to the
reduced descriptions corresponding to regions (1),...,(5).
We have:

\begin{itemize}

\item The matching conditions for the reduced descriptions of the front
of the pulse and the leading edge at  $T=T_0$ are:
\begin{eqnarray}\begin{array}{l}
v^*-v_k^{(II)}({T-T_1\over \epsilon}) \ll 1, \;
m^* - m_k^{(II)}({T-T_1\over \epsilon}) \ll 1, \\
\end{array}\nonumber\end{eqnarray}
if $\epsilon \ll T_0-T \ll 1$. The uniform approximations of the profiles
for $T\leq T_0$ are:
\begin{eqnarray}\begin{array}{ll}
v_k^{unif} =& v_k^{(II)}({T-T_0\over \epsilon}), \quad n_k^{unif} =  n^*,\\
m_k^{unif} =& m_k^{(II)}({T-T_0\over \epsilon}), \quad h_k^{unif} = h^*.
\end{array}\nonumber\end{eqnarray}

\item The matching conditions for the reduced descriptions of the peak
of the pulse and the leading edge at  $T=T_0$ are:
\begin{eqnarray}\begin{array}{l}
\nu^{(3)}_1(n_k^{(III)}(T),h_k^{(III)}(T))-v_k^{(II)}
({T-T_0\over \epsilon}) \ll 1, \\
m_{\infty}\big(\nu^{(3)}_1(n_k^{(III)}(T),h_k^{(III)}(T))\big)-
m_k^{(II)}({T-T_0\over \epsilon}) \ll 1, \\
n_k^{(III)}(T) -n^* \ll 1,\; h_k^{(III)}(T) -h^* \ll 1,
\end{array}\nonumber\end{eqnarray}
if $\epsilon \ll T-T_0 \ll 1$. The uniform approximations of the profiles
in regions (II)-(III) are:
\begin{eqnarray}\begin{array}{ll}
v_k^{unif} =& v_k^{(II)}({T-T_0\over \epsilon}) + \nu^{(3)}_1
(n_k^{(III)}(T),h_k^{(III)}(T))\\
& - \nu^{(3)}_1(n^*,h^*), \\
m_k^{unif} =& m_k^{(II)}({T-T_0\over \epsilon}) + m_{\infty}\big(
\nu^{(3)}_1(n_k^{(III)}(T),h_k^{(III)}(T))\big)  \\
& - m_{\infty}\big(\nu^{(3)}_1(n^*,h^*)\big), \\
n_k^{unif}=&n_k^{(III)}(T),\; h_k^{unif}=h_k^{(III)}(T).
\end{array}\nonumber\end{eqnarray}

\item The matching conditions for the reduced descriptions of the peak
of the pulse and the trailing edge at  $T=T_1$ are:
\begin{eqnarray}\begin{array}{l}
\nu^{(3)}_1(n_k^{(III)}(T),h_k^{(III)}(T))-v_k^{(IV)}
({T-T_1\over \epsilon}) \ll 1, \\
m_{\infty}\big(\nu^{(3)}_1(n_k^{(III)}(T),h_k^{(III)}(T))\big)-
m_k^{(IV)}({T-T_1\over \epsilon}) \ll 1, \\
n_k^{(II)}(T) -n^* \ll 1,\; h_k^{(II)}(T) -h^* \ll 1,
\end{array}\nonumber\end{eqnarray}
if $\epsilon \ll T_1-T \ll 1$. The uniform approximations of the profiles
in regions (II)-(III)-(IV) are:
\begin{eqnarray}\begin{array}{ll}
v_k^{unif} =& v_k^{(II)}({T-T_0\over \epsilon}) +
v_k^{(IV)}({T-T_1\over \epsilon})  \\
& + \nu^{(3)}_1(n_k^{(III)}(T),h_k^{(III)}(T))\\
& - \nu^{(3)}_1(n^*,h^*) -\nu^{(3)}_1(n^{[1]},h^{[1]}), \\
m_k^{unif} =& m_k^{(II)}({T-T_0\over \epsilon}) +
m_k^{(IV)}({T-T_1\over \epsilon}) \\
&+ m_{\infty}\big(\nu^{(3)}_1(n_k^{(III)}(T),h_k^{(III)}(T))\big)  \\
& - m_{\infty}\big(\nu^{(3)}_1(n^*,h^*)\big)
- m_{\infty}\big(\nu^{(3)}_1(n^{[1]},h^{[1]})\big), \\
n_k^{unif}=&n_k^{(III)}(T),\; h_k^{unif}=h_k^{(III)}(T).
\end{array}\nonumber\end{eqnarray}

\item The matching conditions for the reduced descriptions of the tail
of the pulse and the trailing edge at  $T=T_1$ are:
\begin{eqnarray}\begin{array}{l}
\nu^{(1)}_1(n_k^{(V)}(T),h_k^{(V)}(T))-v_k^{(IV)}
({T-T_1\over \epsilon}) \ll 1, \\
m_{\infty}\big(\nu^{(1)}_1(n_k^{(V)}(T),h_k^{(V)}(T))\big)-
m_k^{(IV)}({T-T_1\over \epsilon}) \ll 1, \\
n_k^{(V)}(T) -n^* \ll 1,\; h_k^{(V)}(T) -h^* \ll 1,
\end{array}\nonumber\end{eqnarray}
if $\epsilon \ll T-T_1 \ll 1$. The uniform approximations of the profiles
for $T\geq T_1$ are:
\begin{eqnarray}\begin{array}{ll}
v_k^{unif} =& v_k^{(IV)}({T-T_1\over \epsilon}) + \nu^{(1)}_1
(n_k^{(V)}(T),h_k^{(V)}(T))\\
& - \nu^{(1)}_1(n^{[1]},h^{[1]}), \\
m_k^{unif} =& m_k^{(IV)}({T-T_1\over \epsilon}) + m_{\infty}\big(
\nu^{(1)}_1(n_k^{(V)}(T),h_k^{(V)}(T))\big)  \\
& - m_{\infty}\big(\nu^{(1)}_1(n^{[1]},h^{[1]})\big), \\
n_k^{unif}=&n_k^{(V)}(T),\; h_k^{unif}=h_k^{(V)}(T).
\end{array}\nonumber\end{eqnarray}

\end{itemize}

\section{Pulses formed by two wavefronts}

\label{sec:twointerfaces}

For the choice of parameters indicated in (\ref{frog}), the third
region of the pulse ends when two branches of roots of the cubic
source collapse. For other parameters, the situation might be
different. It might happen that along the integral curve (\ref{w1})
we find a couple $(n^{[1]},h^{[1]})$ such that there are wave front
solutions of the reduced fast system traveling with speed
$c_-(n^{[1]},h^{[1]})=c_+(n^*,h^*)$. Then, the top of the pulse ends
at this point and the fourth region is now a traveling wave front.
The pulse is formed by two rigidly moving traveling wave fronts.

Whether a traveling wave front is formed in the back of the pulse or
not, can be guessed from the shape of the nonlinear source $f$.
Figure \ref{figura1}(a) depicts  $f(v;n,h)$  with the parameters
values (\ref{frog}) when $(n,h)=(n^*,h^*)$. It is strongly
asymmetric. The magnitude of the speed of the leading wave front is
intuitively related to size of the area enclosed by $f(v;n^*,h^*)$
between its second and  third zeroes, $A^*$. If we vary $(n,h)$
along the curve (\ref{w1}), the speed of a back front is related to
the area $A$ enclosed by $f(v;n,h)$ between its first and the second
zeroes. We find that such areas are always smaller than $A^*$ and
the cubic structure is finally lost. A trailing wave front moving at
the same speed as the leading wave front cannot be formed. Back wave
fronts can only be observed for more symmetrical sources, when
varying $(n,h)$ along the curve (\ref{w1}) we can make
 $A$ equal to $A^*$.

We describe here how to modify the asymptotic construction in Sections
\ref{sec:pulses} and \ref{sec:wavetrains} to account for pulses formed
by  two rigidly moving wave fronts.
In the asymptotic description of these pulses we distinguish again five
regions. The first three are similar:

\begin{itemize}
\item The front of the pulse  is described by  $v_k\sim v^*$,
$m_k\sim m^*$, $n_k\sim n^*$ and $h_k\sim h^*$.
\item The leading edge of the pulse is a wave front solution
of the fast reduced system (\ref{f1})-(\ref{f2}) with $n_k=n^*$
and $h_k=h^*$ joining
$(\nu^{(1)}(n^*,h^*),m_{\infty}(v^*)) =(v^*,m^*)$ and
$(\nu^{(3)}(n^*,h^*),m_{\infty}(\nu^{(3)}(n^*,h^*))$.
This front propagates with a definite speed $c=c_+(n^*,h^*)$.
\item In the transition between interfaces, $v_k=\nu^{(3)}(n_k,h_k)$,
 $m_k=m_{\infty}(v_k)$ and $n_k,h_k$ solve the slow reduced
system (\ref{s1}), evolving from $(n^*,h^*)$ to $(n^{[1]},h^{[1]})$.
\end{itemize}

The fourth region is different:

\begin{itemize}
\item The trailing edge is the wave front solution for
the fast reduced system (\ref{f1})-(\ref{f2}) with $n_k=n^{[1]}$
and $h_k=h^{[1]}$, joining
$(\nu^{(3)}(n^{[1]},h^{[1]}),m_{\infty}(\nu^{(3)}(n^{[1]},h^{[1]}))$ and
$(\nu^{(1)}(n^{[1]},h^{[1]}),m_{\infty}(\nu^{(1)}(n^{[1]},h^{[1]}))$,
respectively.  $n^{[1]}$ and $h^{[1]}$ are now selected in
such a way that this front travels with speed
$c=c_-(n^{[1]},h^{[1]})=c_+(n^*,h^*)$.
\end{itemize}

The fifth is similar:

\begin{itemize}
\item In the pulse tail, $v_k=\nu^{(1)}(n_k,h_k)$,
$m_k=m_{\infty}(v_k)$ and $n_k,h_k$ solve the slow reduced
system (\ref{s1}),  evolving from $(n^{[1]},h^{[1]})$ to $(n^*,h^*)$
as $t \rightarrow \infty$.
\end{itemize}

The temporal and spatial width of the peak can be computed as in
Section \ref{sec:pulses}. Now, the number of points in the
leading and trailing wave fronts are neglected.

For wave trains, the construction in Section \ref{sec:wavetrains}
has to be modified as follows: $(N,H)$ and $(n^{[1]},h^{[1]})$ must
lie in the same integral curve of (\ref{wt1}) and satisfy
$c=c_-(n^{[1]},h^{[1]})=c_+(N,H)$.


\begin{thebibliography}{9}
\bibitem{nota} In fact, stages (3)-(5) can be
considered as one region described by the reduced system of four
ordinary differential equations obtained setting $D=0$. The three
regions arise by further reducing the system using the splitting
between fast and slow variables.
\end{thebibliography}
\end{document}